# Schizophrenia-mimicking layers outperform conventional neural network layers

**Ryuta Mizutani[1*], Senta Noguchi[1], Rino Saiga[1], Yuichi Yamashita[2], Mitsuhiro Miyashita[3], Makoto Arai[3], Masanari Itokawa[3,4]**

[1]Department of Applied Biochemistry, Tokai University, Hiratsuka, Kanagawa 259-1292, Japan

[2]Department of Information Medicine, National Institute of Neuroscience, National Center of Neurology and Psychiatry, Kodaira, Tokyo 187-8502, Japan

[3]Tokyo Metropolitan Institute of Medical Science, Setagaya, Tokyo 156-8506, Japan

[4]Tokyo Metropolitan Matsuzawa Hospital, Setagaya, Tokyo 156-0057, Japan

* Correspondence:
Ryuta Mizutani
mizutanilaboratory@gmail.com

## Abstract

We have reported nanometer-scale three-dimensional studies of brain networks of schizophrenia cases and found that their neurites are thin and tortuous compared to healthy controls. This suggests that connections between distal neurons are suppressed in microcircuits of schizophrenia cases. In this study, we applied these biological findings to the design of schizophrenia-mimicking artificial neural network to simulate the observed connection alteration in the disorder. Neural networks having a "schizophrenia connection layer" in place of a fully connected layer were subjected to image classification tasks using the MNIST and CIFAR-10 datasets. The results revealed that the schizophrenia connection layer is tolerant to overfitting and outperforms a fully connected layer. The outperformance was observed only for networks using band matrices as weight windows, indicating that the shape of the weight matrix is relevant to the network performance. A schizophrenia convolution layer was also tested using the VGG configuration, showing that 60% of the kernel weights of the last three convolution layers can be eliminated without loss of accuracy. The schizophrenia layers can be used instead of conventional layers without any change in the network configuration and training procedures; hence, neural networks can easily take advantage of these layers. The results of this study suggest that the connection alteration found in schizophrenia is not a burden to the brain, but has functional roles in brain performance.

## 1    Introduction

Artificial neural networks were originally designed by modelling the information processing of the brain (Rosenblatt, 1958). The primate brain is subdivided into functionally different areas, such as the visual cortex of the occipital lobe and auditory cortex of the temporary lobe (Brodmann, 1909; Amunts & Zilles, 2015). Studies on the visual cortex (Hubel & Wiesel, 1959) inspired the development of the convolutional neural network (Fukushima, 1980) that has evolved into a wide variety of network configurations (Simonyan & Zisserman, 2014; He et al., 2016). Structural analysis of human brain networks and incorporation of resultant biological knowledge into artificial intelligence algorithms have the potential to improve the performance of machine learning.

Analysis of not only healthy control cases but also cases with psychiatric disorders can provide clues to the design of new artificial neural networks. It has been reported that polygenic risk scores for schizophrenia and bipolar disorder were associated with membership in artistic societies and creative professions (Power et al., 2015). A higher incidence of psychiatric disorders was found in geniuses and their families than in the average population (Juda, 1949). This suggests that a possible distinguishing feature of neuronal networks of the psychiatric cases can be exploited in the design of unconventional architectures for artificial intelligence.

We recently reported nanometer-scale three-dimensional studies of neuronal network of schizophrenia cases and age/gender-matched controls by using synchrotron radiation nanotomography or nano-CT (Mizutani et al., 2019, 2021). The results indicated that the neurites of the schizophrenia cases are thin and tortuous, while those of the control cases are thick and straight. The nano-CT analysis also revealed that the diameters of the neurites are proportional to the diameters of the dendritic spines which form synaptic connections between neurons. It has been reported that thinning of neurites or spines attenuates firing efficiency of neurons (Spruston, 2008) and hence affects the activity of the areas to which they belong.

In this study, we incorporated these biological findings in artificial neural networks to delineate 1) how well the neuronal microcircuit tolerate the structural alterations observed in schizophrenia and 2) how we can incorporate those findings into an artificial neural network to improve its performance. The analyses were performed by using newly designed layers that mimic the connection alteration in schizophrenia. The obtained results indicated that the schizophrenia layers tolerate parameter reductions up to 80% of the weights and outperform conventional layers.

## 2    Material and Methods

### 2.1    Design of schizophrenia-mimicking layers

The etiology of schizophrenia has been discussed from neurodegenerative and neurodevelopmental standpoints (Allin & Murray, 2002; Gupta & Kulhara, 2010). The neurodegenerative hypothesis claims that schizophrenia is a disorder due to degeneration in the brain. Another neurodevelopmental hypothesis proposes that the brain network forms abnormally during the developmental process. The etiology of schizophrenia has also been discussed on the basis of the two-hit hypothesis (Maynard et al., 2001), wherein the "first hit" during early development primes the pathogenic response and "second hit" later in the life causes the disorder.

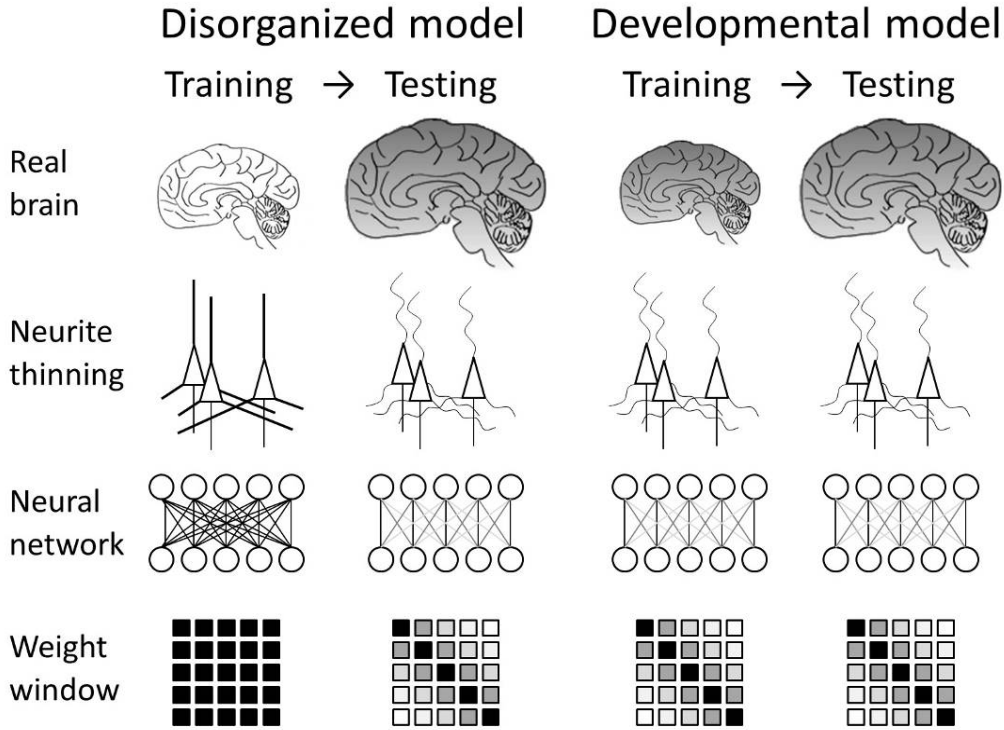

**Figure 1.** Neural network models mimicking schizophrenia. Our nanotomography study revealed that neurites become thin and tortuous in schizophrenia. We translated this finding into two models. In the disorganized model, the structural alteration is incorporated as neurodegeneration after the formation of the brain network. This can be simulated by training the network with the usual manner and then testing it while damping weights between distal nodes. The developmental model assumes concurrent progress of neuropathological changes and brain development. This can be simulated by using a connection-modified layer, in which distal connections are damped during both training and testing.

We translated these understandings of the disorder into two working models of artificial neural network (Figure 1). The first one is a disorganized model. This model mimics neurodegeneration after the formation of the cerebral neuronal network. This can be simulated by training an artificial neural network in the usual manner and then disorganizing it, so that the *posteriori* disorganization simulates neurodegeneration after the formation of the network. Another working model is the developmental model, in which we assume concurrent progress of neuropathological changes and brain development. This developmental model can be simulated by implementing a connection-modified layer in the neural network, which is trained under the modification. Analysis of these models should reveal how the intervention affects network performance.

The thinning of neurites in schizophrenia (Mizutani et al., 2019, 2021) should hinder transmission of the input potentials depending on the length of the neurite from the soma (Spruston, 2008). Therefore, distal synaptic connections should deteriorate more than proximal connections. This phenomenon can be reproduced in an artificial neural network by defining a distance measure between the nodes and by damping the connection parameters depending on the distance measure. Here, we assume a one-dimensional arrangement of nodes and define the distance $d_{ij}$ of the connection between input node $i_x$ and output node $j_y$ as $d_{ij} = |r \cdot i_x - j_y|/\sqrt{r^2 + 1}$, where $r = n_y/n_x$ is the ratio of the number of nodes between the target and the preceding layers. This distance measure is equal to the Euclidean distance between an off-diagonal element and the diagonal in the weight matrix. Since this distance measure is defined in terms of the number of nodes, it is proportional to the neuronal soma size (typically 10–30 µm), and hence can be approximately converted into a real distance by multiplying it by the soma size. The window matrix was prepared by using the above distance measure to modify the weight matrix. Figure 2 shows examples of window matrixes having identical numbers of inputs and outputs. Diagonal connection alteration (Figure 2B) is performed by zeroing the weight parameters if their distances from the diagonal are larger than a threshold. This can be implemented by masking the weight matrix with a window matrix $F = (f_{ij})$, where elements $f_{ij}$ distal to the diagonal are set to 0 and elements $f_{ij}$ proximal to the diagonal are set to 1. A Gaussian window (Figure 2C) has matrix elements $f_{ij}$ of a Gaussian form: $f_{ij} = \exp(-d_{ij}^2/2\sigma^2)$, where $\sigma$ represents the window width. Other window variations (Figure 2D–F) designed differently from the above-mentioned distance idea were also used (Figure 2). The parameter reduction ratio was defined as the ratio between the sum of window elements $\sum_{ij} f_{ij}$ and the total number of weights. The weight matrix was multiplied by the window matrix in an element-by-element manner and then normalized with the parameter reduction ratio so as to keep the sum of weights unchanged.

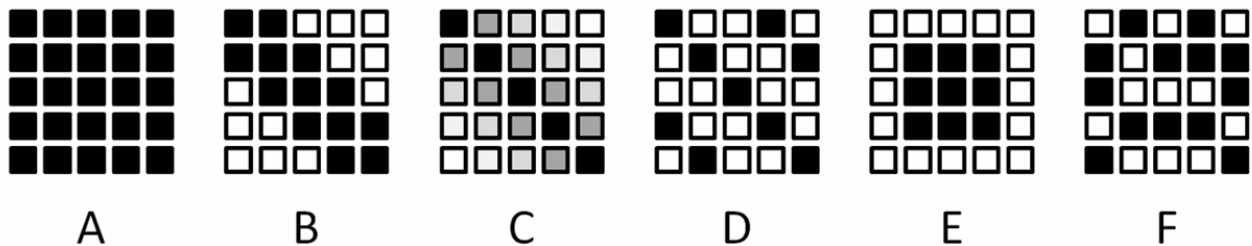

**Figure 2.** Schematic drawings of weight windows. These drawings assume a 5 × 5 square matrix of the weight. Each tiny box indicates an element of the matrix. Window values are represented with gray scale from 0 (white) to 1 (black). (A) Fully connected network. (B) Diagonal window representing connection alteration in schizophrenia. Weight elements indicated with black boxes were used for training and/or evaluation. Elements of open boxes were set to zero, and were not used in the training and/or evaluation. (C) Gaussian window mimicking distance-dependent gradual decrease in the real neuronal connection. (D) Stripe window. (E) Centered window. (F) Random window.

## 2.2 Implementation and examination of schizophrenia layers

The influences of these connection alterations on the neural network were examined using the MNIST handwritten digits dataset (LeCun et al., 1998) and the CIFAR-10 picture dataset (Krizhevsky, 2009). Hereafter, we call the fully connected layer masked with the schizophrenia window the "schizophrenia connection layer" and the convolution layer with the schizophrenia window the "schizophrenia convolution layer". The network configurations used for the image classification tasks are summarized in Table 1 and fully described in Supplementary Table 1. Simple 3- and 4-layer configurations (Table 1, networks A and B) were used in the MNIST classification tasks. Network A with one schizophrenia connection layer as a hidden layer was used for the analysis of the developmental model, in which the connection alteration was incorporated in the training and the evaluation. Network B was used for the analysis of the disorganized model, in which the connection alteration was incorporated only in the evaluation step. In network B, a pair of a fully connected layer and a schizophrenia connection layer with identical numbers of nodes was implemented as hidden layers to prepare square weight matrixes of different sizes, which were used to analyze the effect of the dimension of the weight matrix on the connection alteration. Convolutional networks C–E (Table 1) were used in the classification tasks run on the CIFAR-10 dataset. The configuration of networks C and D was taken from the Keras example code. These networks were used for testing the schizophrenia connection layer as top layers. Network C was used for the analysis of the developmental model and D for the disorganized model. Network E was used for testing the schizophrenia convolution layer along with the schizophrenia connection layer in the VGG16 configuration (Simonyan & Zisserman, 2014). The elements of the kernels of the convolution layers can be regarded as two-dimensional weight arrays, which were masked with the diagonal window. Batch normalization (Ioffe & Szegedy, 2015) and dropout (Srivastava et al., 2014) layers were also incorporated in network E (Supplementary Table 1E).

**Table 1.** Network configuration. Numbers in parentheses represent the number of nodes or number of filters. Further information are shown in Supplementary Table 1. Sz, schizophrenia connection layer; FC, fully connected layer; FC>Sz, trained as a fully connected layer and evaluated using the schizophrenia window; Conv, 2-dimensional convolution layer; VGG16Conv3, the first 3 convolutional blocks of the VGG16 network; SzConv, 2-dimensional schizophrenia convolution layer. A kernel size of 3 × 3 was used for all convolution layers. *Dimensions of these hidden layers were set equal to each other and varied to analyze its effect on the connection alteration.

| A | B | C | D | E |
|---|---|---|---|---|
| Developmental model | Disorganized model | Developmental model | Disorganized model | Developmental model |
| Input (28 × 28) | Input (28 × 28) | Input (32 × 32 RGB) | Input (32 × 32 RGB) | Input (32 × 32 RGB) |
| Sz (512) | FC (64–1024)* | Conv (32) | Conv (32) | VGG16Conv3 |
| Output (10) | FC>Sz (64–1024)* | Conv (32) | Conv (32) | SzConv (512) |
| | Output (10) | Maxpool | Maxpool | SzConv (512) |
| | | Conv (64) | Conv (64) | SzConv (512) |
| | | Conv (64) | Conv (64) | Maxpool |
| | | Maxpool | Maxpool | FC or Sz (4096) |
| | | FC or Sz (512) | FC>Sz (512) | FC or Sz (4096) |
| | | Output (10) | Output (10) | FC or Sz (1024) |
| | | | | Output (10) |

### 2.3    Computational experiments

Training and evaluation of networks A-D were conducted using Tensorflow 2.3.0 and Keras 2.4.0 running on the c5a.xlarge (4 vCPUs of AMD EPYC processors operated at 2.8 GHz) or the c5a.2xlarge (8 vCPUs) instance of Amazon Web Service. Training and evaluation of network E were conducted using Tensorflow 2.7.0 and Keras 2.7.0 running on the same instances. The CPU time required for training and evaluating the networks using the schizophrenia layers was slightly shorter than those using the normal layers, though the incorporation of the Gaussian window required additional time to initialize its window elements. The Python codes used in this study are available from our GitHub repository (https://mizutanilab.github.io). Statistical analyses were conducted using R 3.4.3. Significance was defined as $p < 0.05$.

Biases were enabled in all layers, except for the schizophrenia layers in the disorganized model. This is because biases can be refined in the developmental model but cannot be modified according to the inter-node distance in the evaluation step. The ReLU activation function (Glorot et al., 2011) was used in all of the hidden layers, while softmax was used in the output layers. Hidden layers were initialized with He's method (He et al., 2015). Networks A, B and E were trained using the Adam algorithm (Kingma & Ba, 2014). Networks A and B were trained with a learning rate of $1 \times 10^{-3}$. Network E was trained with a learning rate of $5 \times 10^{-4}$ first, and then with $1 \times 10^{-4}$ after 150 epochs. Networks C and D were trained using the RMSprop algorithm (Tieleman & Hinton, 2012) with a learning rate of $1 \times 10^{-4}$ and decay of $1 \times 10^{-6}$. Batch sizes were set to 32 for networks A–D, and 200 for network E. Data augmentation (Wong et al., 2016) was used in the training of network E.

## 3    Results

### 3.1    Disorganized models

Figure 3 summarizes the relationships between the connection alteration and the classification error in the disorganized model, in which training precedes the alteration. Figure 3A shows the dependence on window shape in the MNIST classification tasks, which were conducted using a 4-layer network (Table 1, B). Weight parameters between two hidden layers of identical numbers of nodes were modified after training in order to mimic neurodegeneration after formation of the neuronal network. The resultant modified network was subjected to the evaluation using the validation dataset. The results indicated that network B can tolerate a parameter reduction of up to approximately 60% of the connections between the hidden layers (Figure 3A). The profiles showed little dependence on the window shape except for the centered window, indicating that the contribution of each weight element to the network performance is equivalent independently of their position in the weight matrix.

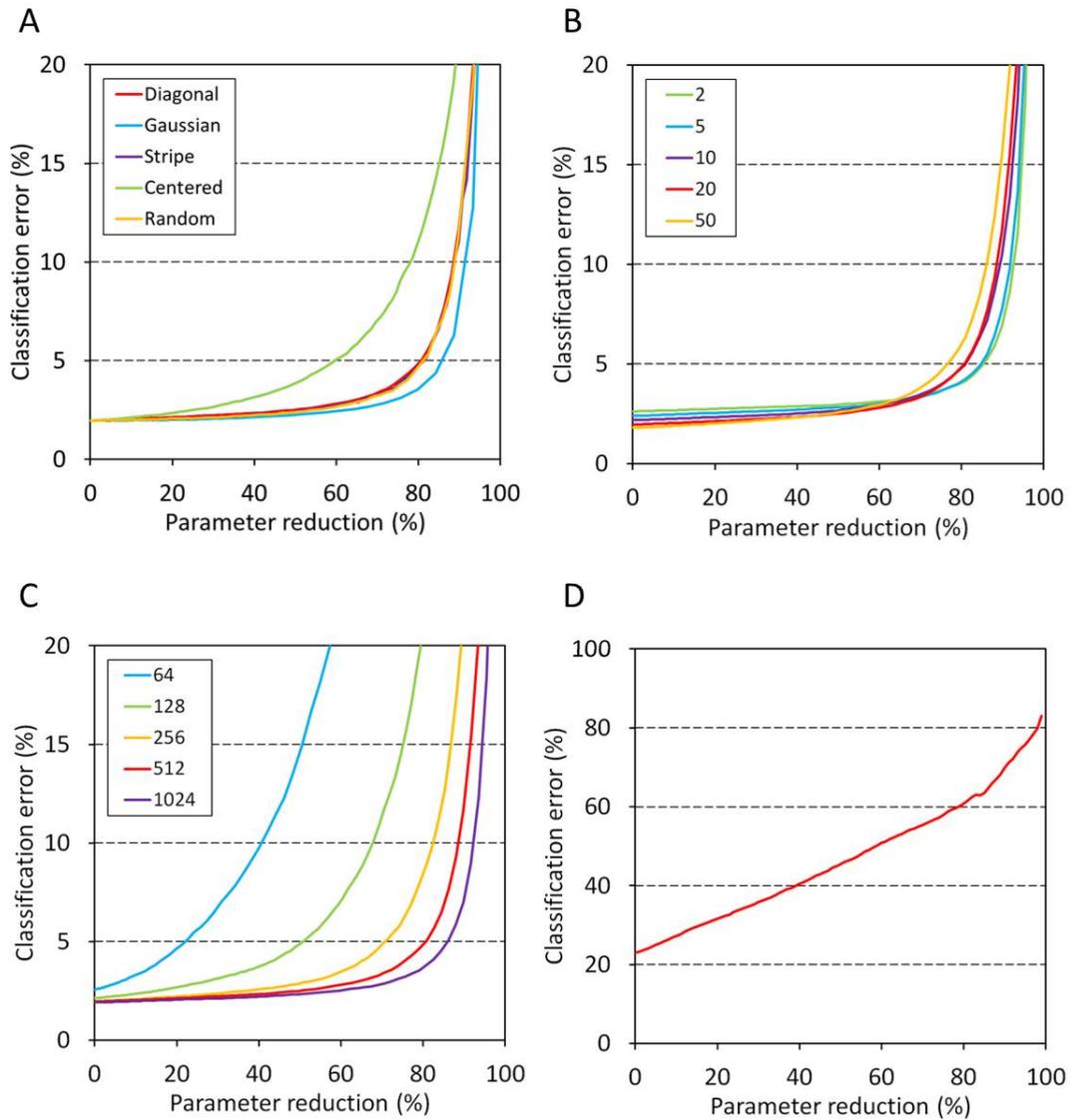

**Figure 3.** Relationship between classification error and parameter reduction in the disorganized model. Networks were trained first and then the weights of the target layers were masked with a window in the evaluation. The left intercept corresponds to a network with fully connected layers. Panels A–C show the results of MNIST classification. Unless otherwise stated, network B (Table 1) having 512 nodes in the two hidden layers was trained for 20 epochs and then its schizophrenia connection layer was masked using a diagonal window in the evaluation. The training and evaluation were repeated for 100 sessions and their mean errors were plotted. (A) Effect of window shape. (B) Dependence on training duration. (C) Dependence on hidden layer dimensions. (D) Results of CIFAR-10 classification. Network D (Table 1) was trained for 100 epochs, and its top layer was masked using a diagonal window. The training and evaluation were repeated for 30 sessions, and their mean errors are plotted.

Figure 3B shows the dependence on the number of epochs of the disorganized model on the MNIST tasks. The results indicated that the network became slightly more sensitive as the training became longer, suggesting that the redundancy of the weight matrix elements decreased after a long training duration. Figure 3C shows the relation between the number of nodes and tolerance against connection alteration. The networks having 64 or 128 nodes in the hidden layers became prone to error due to the parameter reduction, while networks having 256 or more nodes in the hidden layers tolerated a parameter reduction up to 60–80% of the weights. These results indicate that the networks having 256 nodes or more have sufficient parameters to store the trained information. In contrast, the CIFAR-10 classification task using network D (Table 1) showed an increase in error that was nearly proportional to the parameter reduction (Figure 3D). This indicates that the information acquired during training is uniformly but not redundantly distributed in the top layer, resulting in the low tolerance of the network against the parameter reduction.

## 3.2    Developmental models

The developmental model showed distinct features that were not observed in the disorganized models. Figure 4A shows the progress of training the developmental model (Table 1, C) on the CIFAR-10 classification tasks, in which the weights of the top layer were masked with a diagonal window throughout the training and evaluation. The task was performed using network C, which consisted of two blocks of convolution layers and one schizophrenia top layer. A network with the same configuration but having a fully connected top layer was used as a control. The obtained results revealed that the schizophrenia network outperformed the control network. The control network showed overfitting approximately after 75 epochs of training, whereas the schizophrenia network showed a continuous decline in error out to 200 epochs. The classification error of the schizophrenia network was significantly lower than that of the control even before the overfitting ($p = 0.014$ at 75 epochs, and $p = 1.1 \times 10^{-5}$ at 200 epochs, two-sided Wilcoxon test, $n_1 = n_2 = 10$). The overfitting of the control network was not suppressed by using the dropout method.

The connection alteration was also incorporated in the convolution layers by masking the kernel elements with the diagonal window. This schizophrenia convolution layer was implemented in the VGG16 configuration (Table 1, E) to perform the CIFAR-10 classification tasks. Figure 4B shows the progress of training. The schizophrenia network with a 60% parameter reduction in the last three convolution layers performed comparably to the control network. This result indicates that the convolution layers of this network contain parameter redundancy that can be eliminated by using the diagonal window. We further replaced the top layers of the VGG network with schizophrenia connection layers and examined the network's performance of the same CIFAR-10 classification task. The obtained results (Supplementary Figure 1) indicated that half the top layer weights of the VGG network can be eliminated without loss of accuracy by using the schizophrenia-connection top layers.

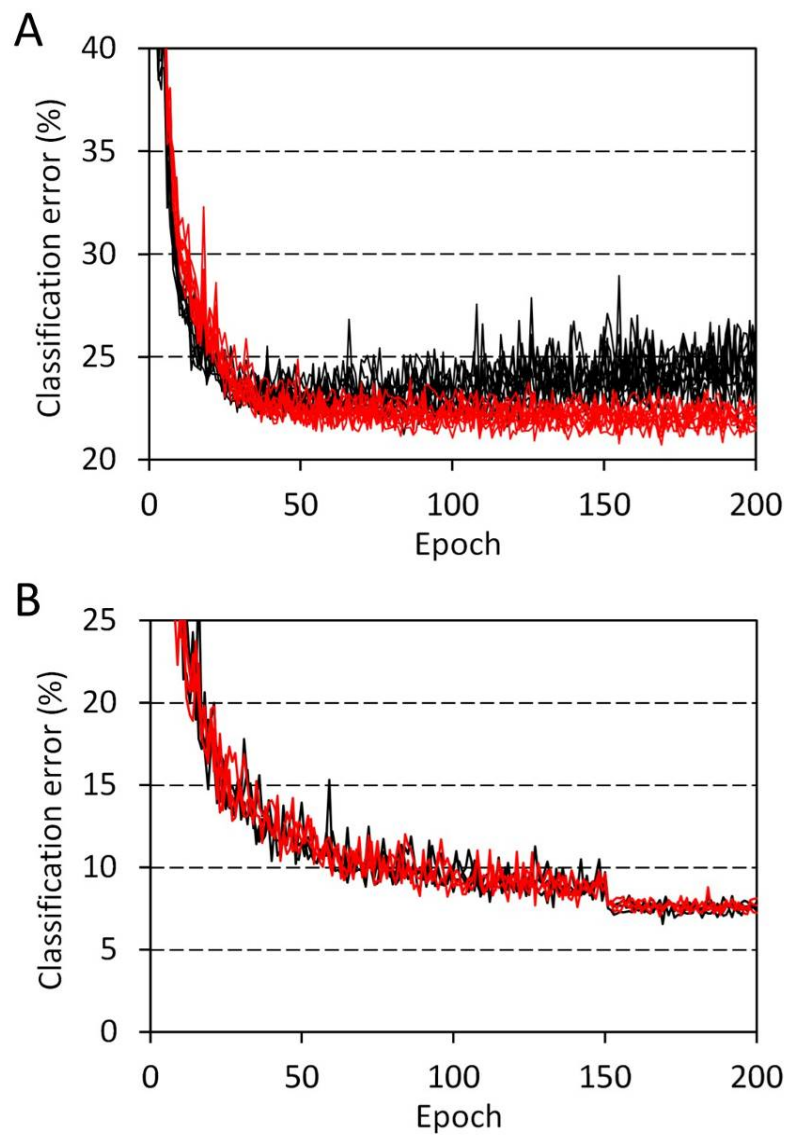

**Figure 4.** Progress of training in the CIFAR-10 classification tasks. (A) Network C (Table 1) with a schizophrenia connection layer was trained for 10 sessions, and the resultant errors are plotted in red. Parameter reduction in the schizophrenia layer was set to 50%. Errors of the control network with the same configuration but having a fully connected layer are plotted in black. (B) VGG network E (Table 1) with schizophrenia convolution layers was trained for 3 sessions, and the resultant errors are plotted. Results for the schizophrenia network with a 60% parameter reduction in the last three convolution layers are drawn in red, while those of a control network without the parameter reduction are drawn in black.

In order to visualize the response of the learned filters of the schizophrenia convolution layer, we replaced the first convolution layer of the VGG16 network (Table 1, E) with a schizophrenia convolution layer and performed the same CIFAR-10 classification task. Over 40% of the weights of the first convolution layer were eliminated without loss of accuracy, as shown in Supplementary Figure 2. The obtained responses of the learned filters are shown in Figure 5. These results illustrate that the learned filters of the schizophrenia convolution layer decompose image inputs into RGB channels. This is because the distance in the weight matrix of the schizophrenia convolution layer is defined along the channel dimensions so that the convolutional filters can decode information in a channel-wise manner. In contrast, the filters of the conventional convolution layer showed color-independent patterns due to the absence of restrictions on the weight matrix. Although each kernel of the schizophrenia convolution layer is mostly composed of three primary colors, the network accuracy was the same as that of the native VGG16 network (Supplementary Figure 2), indicating that the RGB decomposition in the first convolution layer imposes no limitation on image recognition.

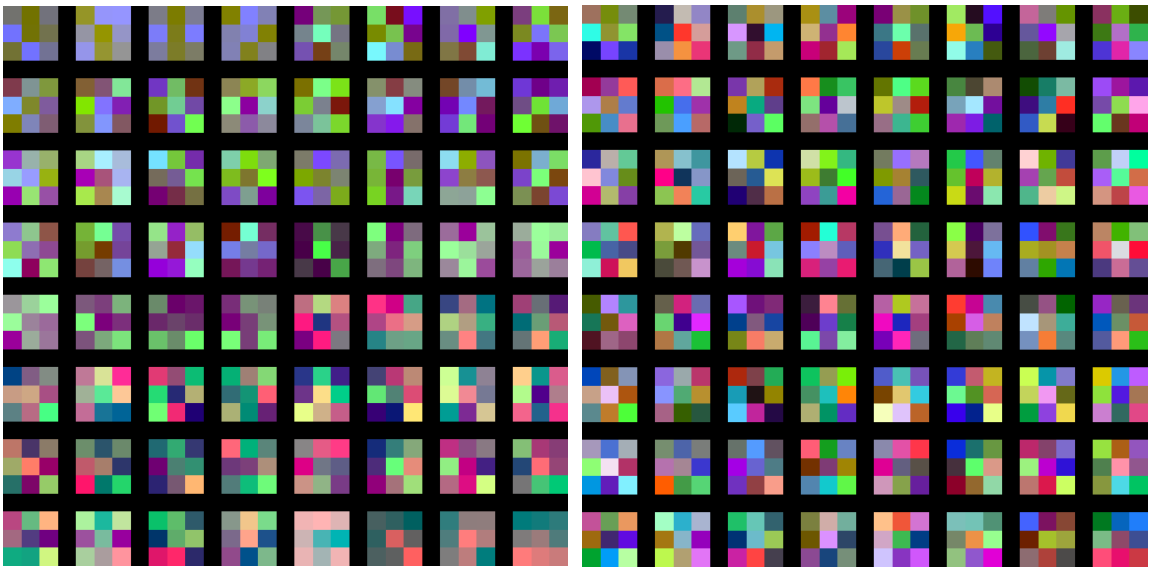

**Figure 5.** Learned filters of (A) schizophrenia convolution layer and (B) conventional convolution layer. In the schizophrenia network, the first convolution layer of the VGG16 network (Table 1, E) was replaced with a schizophrenia convolution layer, in which 41.7% of the weights were eliminated with a diagonal window.

The relation between performance and the parameter reduction ratio in the developmental model was also examined. Figure 6A shows the results of the three-layer networks on the MNIST classification tasks (Table 1, A). The classification error of the schizophrenia network gradually decreased to below that of the control network as the parameter reduction was increased to 70%. The profiles obtained using diagonal and Gaussian windows were almost the same, though the Gaussian window showed stronger tolerance to the parameter reduction than the diagonal window. The stronger tolerance of the Gaussian window suggests that the bilateral tails of the Gaussian function allowed weak connections between distal nodes and mitigated weight masking with the window. In contrast, the network using a random window showed no decrease in error (Figure 6A), indicating that the shape of the diagonal or Gaussian window is relevant to the performance. Figure 6B shows the results of network C on the CIFAR-10 classification tasks. The relation between the error and the parameter reduction was similar to that observed in the MNIST tasks. The profiles shifted toward the lower right as the duration of training became longer, indicating that a schizophrenia connection layer with a larger parameter reduction performs better by training longer. We also examined the effect of conventional $L_1$ regularization by using network C (Table 1). We replaced the schizophrenia top layer of network C with a fully connected layer and introduced conventional $L_1$ regularization to that layer. This $L_1$ regularized network showed virtually no improvement in error (Figure 6B), indicating again that the outperformance is ascribable to the schizophrenia layer.

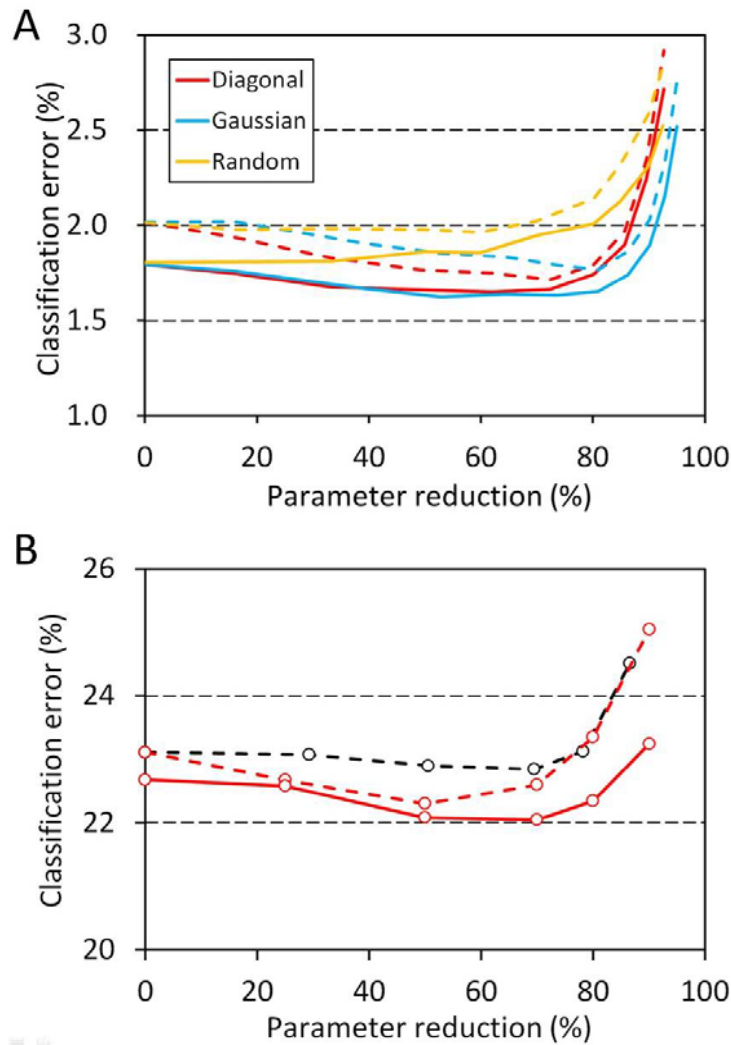

**Figure 6.** Relation between error and parameter reduction in the developmental model, in which the weights of the schizophrenia connection layers were masked with a window throughout the training and evaluation. The left intercepts correspond to a network with fully connected layers. (A) MNIST classification using network A (Table 1). Training and evaluation were repeated for 100 sessions, and their mean classification errors are plotted. Errors at 10 epochs are plotted with dashed lines and errors at 50 epochs with solid lines. (B) CIFAR-10 classification using network C (Table 1). Training and evaluation were repeated for 10 sessions, and their mean errors are plotted. A diagonal window was used in this task. Errors after 50 epochs are plotted with red dashed lines and errors after 75 epochs with red solid lines. The same classification task was also performed with conventional $L_1$ regularization for 10 sessions, and the resultant mean errors after 50 epochs are plotted with black dashed lines. Weights less than 0.0001 were regarded as zero in the reduction ratio calculation.

## 4    Discussion

### 4.1    Related works

Neural networks were first developed by incorporating biological findings, but until now, the structural aspects of neurons of patients with psychiatric disorders have not been incorporated in studies on artificial intelligence. This is probably because the neuropathology of psychiatric disorders had not been three-dimensionally delineated (Itokawa et al., 2020) before our recent reports regarding the nanometer-scale structure of neurons of schizophrenia cases (Mizutani et al., 2019, 2021). A method called "optimal brain damage" (Le Cun et al, 1990) has been proposed to remove unimportant weights to reduce the number of parameters, although its relation to biological findings such as those regarding brain injuries has not been explicitly described.

Parameter reduction and network pruning have been suggested as strategies to simplify the network. It has been reported that simultaneous regularization during training can reduce network connections while maintaining competitive performance (Scardapane et al., 2017). A method to regularize the network structure including the filter shapes and layer depth has been reported to allow the network to learn more compact structures without loss of accuracy (Wen et al., 2016). A study on network pruning has suggested that careful evaluations of the structured pruning method are needed (Liu et al., 2018). Elimination of zero weights after training has been proposed as a way to simplify the network (Yaguchi et al., 2018). Improvements in accuracy have been reported for regularized networks (Scardapane et al., 2017; Yaguchi et al., 2018), although these parameter reduction methods require dedicated algorithms or procedures to remove parameters during training.

Regularization on the basis of filter structure has been reported. The classification accuracy can be improved by using customized filter shapes in the convolution layer (Li et al., 2017). The shape of the filters can also be regularized from the symmetry in the filter matrix (Anselmi et al., 2016). It has been reported that a low-rank approximation can be used to regularize the weights of the convolution layers. (Yu et al., 2017; Idelbayev & Carreira-Perpiñán, 2020). Kernel-wise removal of weights has also been proposed as a regularization method for convolutional networks (Berthelier, 2020). These reports focus on the shape of the image dimensions, while the schizophrenia-mimicking modification of convolution layers proposed in this study is performed by masking the weight matrix with a band matrix defined along the channel dimensions but not along the image dimensions. This strategy allowed us to eliminate 60% of the weights of the last three convolution layers of the VGG16 network without loss of accuracy (Figure 4B). We suggest that the real human brain has already implemented this simple and efficient strategy in the process of its biological evolution.

### 4.2    Schizophrenia-mimicking neural network

We translated recent findings on schizophrenia brain tissue into two schizophrenia-mimicking models: a disorganized model and developmental model (Figure 1). The disorganized model mimics neurodegeneration after the formation of the neuronal network, which can be simulated by training an artificial neural network normally and then damping its weights with the schizophrenia window (Figure 2). The obtained results indicated that the network works even after the post-training intervention, though the alteration did not improve performance  (Figure 3). The developmental model assumes concurrent progress of neuropathological changes and brain development. It was simulated by training and testing the neural network while masking the weight matrix with the schizophrenia window (Figure 2). The results indicated that the schizophrenia connection layer is tolerant to overfitting and outperforms a fully connected layer (Figure 4). Outperformance was only observed in the developmental model and is thus ascribed to the training using the schizophrenia window.

Parameter reduction in schizophrenia layers can be regarded as enforced and predefined $L_1$ regularization. The schizophrenia connection layers using band matrixes as weight windows had the highest levels of performance (Figure 6A). This indicates that the shape of the weight matrix is relevant to the network performance. The convolution layers with a diagonal window performed comparably to the normal convolution layers, revealing that up to 60% of the parameters can be eliminated without a loss of accuracy by using the diagonal window. Training of the schizophrenia network requires no modification of the optimization algorithm, since its parameter reduction is arbitrarily configured *a priori*. Schizophrenia layers can be used instead of conventional layers without any changes in the network configuration. The advantages of schizophrenia layers therefore can be had by any kind of neural network simply by replacing the conventional layers with them.

The structure of the band window matrix of the schizophrenia layer indicates the importance of connecting all nodes, but at the same time, it indicates the importance of dividing them into groups so that each group can process information independently and integratively. The weight window restricts the output nodes to represent only a predefined part of the inputs. Although the random window defines the connections of each node depending on its randomness, the diagonal or the Gaussian window forces the output nodes to divvy up all the inputs, so that all of the input information is grouped and processed into the output nodes. The high performance of the diagonal or the Gaussian window is ascribable to this structural feature of the band matrix. The results shown in Figure 6 indicate that the performance optimum of the schizophrenia layer using the band matrix window is situated nearer to the grouping than to the integration. We recommend a 50–70% parameter reduction as a first choice to obtain the best result.

A wide variety of computational models have been reported for schizophrenia (Lanillos et al., 2020). Elimination of working memory connections in recurrent network was proven to improve perceptual ability, while excess elimination causes output hallucinations under the absence of inputs (Hoffman & McGlashan, 1997). Although the present results indicated that the structural alteration of neurites observed in schizophrenia can affect network performance, its relation to schizophrenia symptoms remains to be clarified. Another limitation of this study is that the present analysis used only thousands of nodes per model and cannot represent the brain-wide disconnectivity observed in the diffusion tensor imaging of schizophrenia cases (Son et al., 2015).

The profiles shown in Figure 6 illustrated that the high level of performance of the schizophrenia layer goes hand in hand with the malfunction. The evolutionary process should geared to finding the level of brain performance that maximizes the survivability of our species. The results of this study, along with the known relationship between creativity and psychosis (Power et al., 2015), suggests that the connection alteration during network development is not a burden to the brain, but has functional roles in cortical microcircuit performance. We suggest that the connection alteration found in schizophrenia cases (Mizutani et al., 2019, 2021) is rationally implemented in our brains in the process of evolution.

## 5    Conflict of Interest

MM, MA and MI declare a conflict of interest, being authors of several patents regarding therapeutic use of pyridoxamine for schizophrenia. All other authors declare that the research was conducted in the absence of any commercial or financial relationships that could be construed as a potential conflict of interest.

## 6 Author Contributions

RM designed the study. RM and SN performed the numerical experiments and analyzed the results. RM wrote the manuscript based on discussions with RS regarding the human neuronal network, with YY regarding the computational models of psychiatric disorders, and with MM, MA, and MI regarding psychiatric disorders. RM prepared the figures.

## 7 Funding

This work was supported by Grants-in-Aid for Scientific Research from the Japan Society for the Promotion of Science (nos. 21611009, 25282250, 25610126, 18K07579, and 20H03608), and by the Japan Agency for Medical Research and Development under grant nos. JP18dm0107088, JP19dm0107088, and JP20dm0107088.

## 8 Acknowledgments

# *Supplementary Material*

# Schizophrenia-mimicking layers outperform conventional neural network layers

Ryuta Mizutani, Senta Noguchi, Rino Saiga, Yuichi Yamashita, Mitsuhiro Miyashita, Makoto Arai, and Masanari Itokawa

**Supplementary Table 1. (A)** Configuration of network A. Sz, schizophrenia connection layer. The Number-of-parameters column shows the number of trainable parameters before the parameter reduction.

| Layer | Output size | Number of parameters | Options |
|-------|-------------|----------------------|---------|
| Input | 28 × 28 | | |
| Sz | 512 | 401,920 | Parameter reduction: 0–95% |
| Output | 10 | 5,130 | |

**Supplementary Table 1. (B)** Configuration of network B. FC>Sz, trained as a fully connected layer and evaluated using a weight window. The Number-of-parameters column shows the number of trainable parameters before the parameter reduction. *Dimensions of these hidden layers were set equal to each other and varied to examine the effect on the connection alteration.

| Layer | Output size | Number of parameters | Options |
|-------|-------------|----------------------|---------|
| Input | 28 × 28 | | |
| Fully connected | 64–1024* | 50,240–803,840 | |
| FC>Sz | 64–1024* | 4,096–1,048,576 | Bias vector was disabled. |
| Output | 10 | 650–10,250 | |

**Supplementary Table 1. (C)** Configuration of network C. Sz, schizophrenia connection layer; Conv, 2-dimensional convolution layer with a kernel size of 3 × 3; FC, fully connected layer. The Number-of-parameters column shows the number of trainable parameters before the parameter reduction.

| Layer | Output size | Filter size | Number of parameters | Options |
|---|---|---|---|---|
| Input | 32 × 32 RGB | | | |
| Conv | 32 × 32 × 32 | 32 | 896 | Zero padding |
| Conv | 30 × 30 × 32 | 32 | 9,248 | No padding |
| Maxpool | 15 × 15 × 32 | | | Pooling 2 × 2, dropout: 25% |
| Conv | 15 × 15 × 64 | 64 | 18,496 | Zero padding |
| Conv | 13 × 13 × 64 | 64 | 36,928 | No padding |
| Maxpool | 6 × 6 × 64 (= 2304) | | | Pooling 2 × 2, dropout: 25% |
| Sz or FC | 512 | | 1,180,160 | Parameter reduction: 0–90% $L_1$ regularization |
| Output | 10 | | 5,130 | |

**Supplementary Table 1. (D)** Configuration of network D. FC>Sz, trained as a fully connected layer and evaluated using a weight window. Conv, 2-dimensional convolution layer with a kernel size of 3 × 3. The Number-of-parameters column shows the number of trainable parameters before the parameter reduction.

| Layer | Output size | Filter size | Number of parameters | Options |
|---|---|---|---|---|
| Input | 32 × 32 RGB | | | |
| Conv | 32 × 32 × 32 | 32 | 896 | Zero padding |
| Conv | 30 × 30 × 32 | 32 | 9,248 | No padding |
| Maxpool | 15 × 15 × 32 | | | Pooling 2 × 2, dropout: 25% |
| Conv | 15 × 15 × 64 | 64 | 18,496 | Zero padding |
| Conv | 13 × 13 × 64 | 64 | 36,928 | No padding |
| Maxpool | 6 × 6 × 64 (= 2304) | | | Pooling 2 × 2, dropout: 25% |
| FC>Sz | 512 | | 1,179,648 | Bias vector was disabled. |
| Output | 10 | | 5,130 | |

**Supplementary Table 1. (E)** Configuration of network E. Conv, 2-dimensional convolution layer; SzConv, 2-dimensional schizophrenia convolution layer; FC, fully connected layer. A kernel size of 3 × 3 was used for all convolution layers. The Number-of-parameters column shows the number of trainable parameters before the parameter reduction.

| Layer | Output size | Filter size | Number of parameters | Options |
|---|---|---|---|---|
| Input | 32 × 32 RGB | | | |
| SzConv | 32 × 32 × 64 | 64 | 1,792 | Zero padding, parameter reduction: 0 or 42% |
| Batch norm. | | | 256 | |
| Conv | 32 × 32 × 64 | 64 | 36,928 | Zero padding |
| Maxpool | 16 × 16 × 64 | | | Pooling 2 × 2, dropout: 25% |
| Conv | 16 × 16 × 128 | 128 | 73,856 | Zero padding |
| Batch norm. | | | 512 | |
| Conv | 16 × 16 × 128 | 128 | 147,584 | Zero padding |
| Maxpool | 8 × 8 × 128 | | | Pooling 2 × 2, dropout: 25% |
| Conv | 8 × 8 × 256 | 256 | 295,168 | Zero padding |
| Batch norm. | | | 1,024 | |
| Conv | 8 × 8 × 256 | 256 | 590,080 | Zero padding |
| Batch norm. | | | 1,024 | |
| Conv | 8 × 8 × 256 | 256 | 590,080 | Zero padding |
| Maxpool | 4 × 4 × 256 | | | Pooling 2 × 2 |
| Conv | 4 × 4 × 512 | 512 | 1,180,160 | Zero padding |
| Batch norm. | | | 2,048 | |
| Conv | 4 × 4 × 512 | 512 | 2,359,808 | Zero padding |
| Batch norm. | | | 2,048 | |
| Conv | 4 × 4 × 512 | 512 | 2,359,808 | Zero padding |
| Maxpool | 2 × 2 × 512 | | | Pooling 2 × 2 |
| SzConv | 2 × 2 × 512 | 512 | 2,359,808 | Zero padding, parameter reduction: 0 or 60% |
| Batch norm. | | | 2,048 | |
| SzConv | 2 × 2 × 512 | 512 | 2,359,808 | Zero padding, parameter reduction: 0 or 60% |
| Batch norm. | | | 2,048 | |
| SzConv | 2 × 2 × 512 | 512 | 2,359,808 | Zero padding, parameter reduction: 0 or 60% |
| Maxpool | 1 × 1 × 512 | | | Pooling 2 × 2 |
| FC or Sz | 4096 | | 2,101,248 | Parameter reduction: 0% (FC) or 50% (Sz) |
| FC or Sz | 4096 | | 16,781,312 | Parameter reduction: 0% (FC) or 50% (Sz) |
| FC or Sz | 1024 | | 4,195,328 | Parameter reduction: 0% (FC) or 50% (Sz) |
| Output | 10 | | 10,250 | |

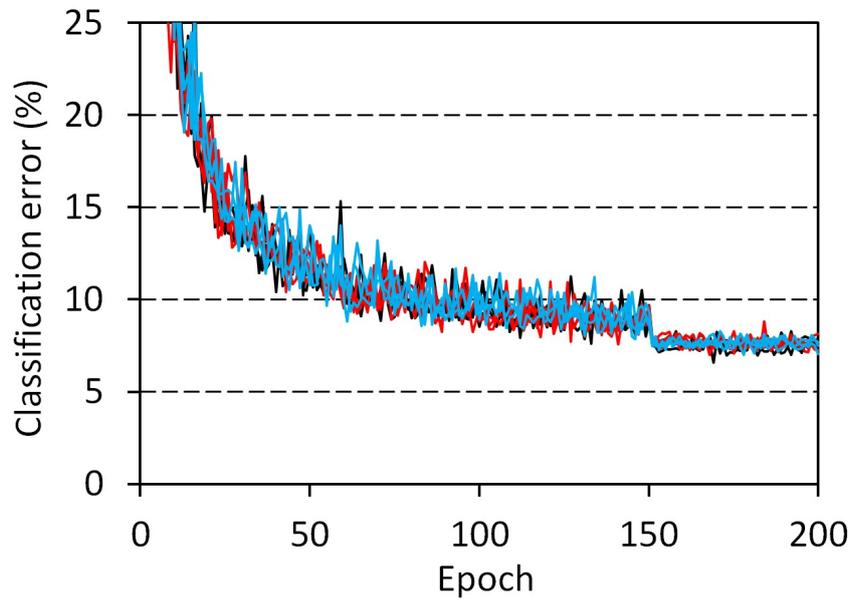

**Supplementary Figure 1.** Progress of training in the CIFAR-10 classification tasks using the VGG16 network. The three top layers were replaced with schizophrenia connection layers in addition to using schizophrenia convolution layers in the last convolutional block (Table 1, E). Parameter reduction was set to 50% in the schizophrenia connection layers and 60% in the schizophrenia convolution layers. This double-schizophrenia VGG16 network was trained for 3 sessions, and the resultant errors are plotted in cyan. Results for the schizophrenia VGG16 network with a 60% parameter reduction only in the last three convolution layers are drawn in red. Those of a control network without schizophrenia layers are drawn in black.

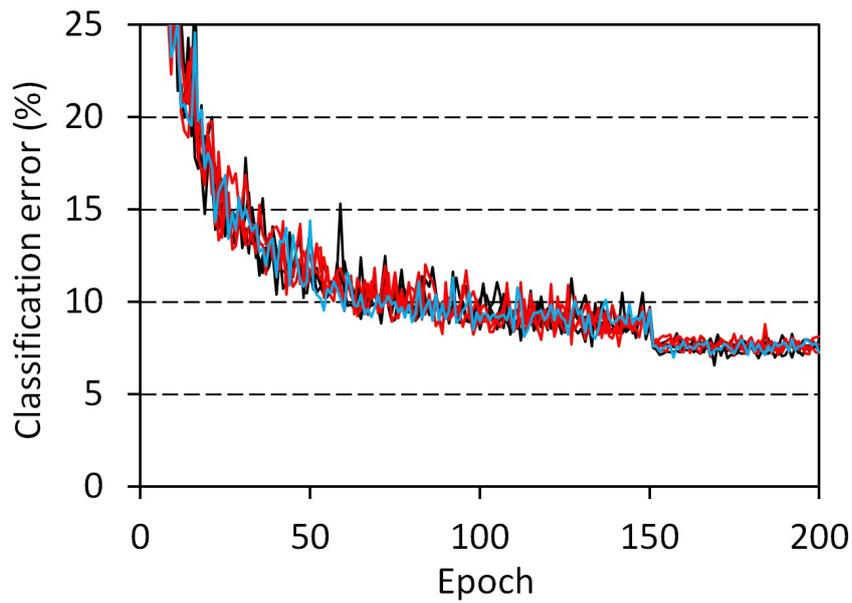

**Supplementary Figure 2.** Progress of training in the CIFAR-10 classification tasks using the VGG16 network. In order to analyze the responses of the learned filters (Figure 5), the first convolution layer was replaced with a schizophrenia convolution layer in addition to using schizophrenia convolution layers in the last convolutional block (Table 1, E). Parameter reduction was set to 41.7% in the first convolution layer and 60% in the last three convolution layers. This schizophrenia VGG16 network was trained for 1 session, and the resultant errors are plotted in cyan. Results for the schizophrenia VGG16 network with a 60% parameter reduction only in the last three convolution layers are drawn in red. Those of a control network without schizophrenia layers are drawn in black.